\documentclass[11pt]{article}

  \usepackage{float}
  \usepackage{placeins}
  \usepackage{acl}
  \usepackage{times}
  \usepackage{latexsym}
  \usepackage[T1]{fontenc}
  \usepackage[utf8]{inputenc}
  \usepackage{microtype}
  \usepackage{inconsolata}
  \usepackage{graphicx}
  \usepackage{booktabs}
  \usepackage{amsmath}
  \usepackage{amssymb}
  \usepackage{dblfloatfix}
  \usepackage{placeins}
  \usepackage{enumitem}
  \setlist[enumerate]{leftmargin=*, topsep=2pt, itemsep=0pt, parsep=0pt, partopsep=0pt}
  \setlist[itemize]{leftmargin=*, topsep=2pt, itemsep=0pt, parsep=0pt, partopsep=0pt}

  \title{NeuroArmor: Safe-Variant-Guided Representation Consistency for Selective Re-Anchoring in Jailbreak Defense}

  \author{
    Zhongyang Lin, Ziran Zhao, Feifei Zhai, Pengyuan Liu \\
  }

\begin{document}
  \maketitle

\begin{abstract}
Large language models remain vulnerable to jailbreak attacks that hide harmful intent behind seemingly ordinary requests such as role-play, translation, encoding, adversarial suffixes, and multi-turn buildup. Existing defenses still struggle to handle these attacks without over-blocking benign but sensitive requests, partly because they often apply the same action to every prompt and therefore fail to balance safety and helpfulness. We propose \textbf{NeuroArmor}, a white-box runtime defense that uses prompt-specific safe variants as a local safety reference for deciding \emph{when} intervention is needed and, once triggered, as safe targets for intervention. For each prompt, NeuroArmor builds $K$ safe variants, compares the prompt state against this local safe reference in hidden-state space, and routes anomalies either to a refusal branch for malicious prompts or to a helpful recovery branch for borderline benign prompts. On Llama-3-8B-Instruct, NeuroArmor reduces malicious attack success rate (\textbf{ASR}) from \textbf{41.56\%} to \textbf{1.57\%} while lowering benign false positive rate (\textbf{FPR}) on the shared benign pool from \textbf{30.26\%} to \textbf{22.05\%}; matched baselines remain substantially weaker on this trade-off. External-judge and manual behavioral evaluations further show that the remaining non-blocked outputs are much less likely to be operationally harmful. Overall, NeuroArmor provides a more effective runtime strategy for jailbreak defense by combining prompt-specific consistency checking, routing, and selective intervention.
\end{abstract}

  \section{Introduction}

  Large language models are increasingly used for question answering, coding, translation, retrieval, and agentic systems~\cite{ouyang2022instructgpt,openai2023gpt4,llama32024herd,gemma22024}. However, the same instruction-following ability that makes them useful also leaves them vulnerable to jailbreak attacks that elicit outputs the model should refuse. In realistic deployments, harmful intent is rarely presented as a direct unsafe request. Instead, it is often hidden behind role-play, hypothetical framing, translation, encoding, adversarial suffixes, or long multi-turn buildup. A practical defense must therefore preserve useful behavior while still maintaining a clear safety boundary.

  Existing defenses can be grouped into three broad families: input filtering, output classification, and training-time alignment. These approaches remain important, but they often act in the same way on every input and offer limited insight into how a jailbreak changes the model internally. This motivates a prompt-specific defense that intervenes only when the input has measurably drifted away from a safe internal reference, while still preserving useful behavior on borderline benign prompts.

Our key observation is that a successful jailbreak often becomes internally inconsistent with a small set of safe variants of the same prompt. Building on this observation, we propose NeuroArmor, a representation-level runtime defense that uses prompt-specific safe variants as a local reference for the model's internal state. This reference serves two roles: it helps identify prompts that have drifted away from the safe region, and it provides prompt-specific safe targets for intervention. The second role matters for borderline benign prompts, because some inputs should be recovered into a helpful, non-harmful interpretation rather than forced into refusal.

  NeuroArmor combines three pieces: prompt-specific safe variants, a consistency detector, and a routing step that decides both \emph{when} to intervene and \emph{which} branch to apply.

  In the reported system, these components are instantiated by a fixed runtime policy that reuses stored variants when available, constructs fallback variants when no stored entry is available, and incorporates lightweight wrapper or multi-turn cues into routing. The evaluation therefore concerns the full prompt-specific routing policy rather than a detector in isolation.

  We make three contributions.

  \begin{enumerate}
  \item We propose a runtime defense in which prompt-specific safe variants serve both as a local reference for consistency checking and as prompt-specific safe targets for intervention.
  \item We introduce a consistency detector and routing rule that use the same prompt-specific evidence to decide both whether to intervene and whether to take a refusal branch or a helpful recovery branch at inference time.
  \item We show through full evaluations and mixed-setting ablations that the gains come from the joint detector-and-routing design, and that the same policy improves both malicious suppression and benign-side calibration in the reported setting.
  \end{enumerate}

  At a high level, NeuroArmor first builds a small prompt-specific set of safe variants, then checks whether the prompt's hidden state remains locally consistent with that set, and finally routes triggered cases either to a refusal branch or to a helpful recovery path. This separation between trigger and action is central: the same safe-variant evidence determines both whether to intervene and where to move the prompt state when intervention is required. On Llama-3-8B-Instruct, the full system lowers malicious success from \textbf{41.56\%} to \textbf{1.57\%} while also reducing pooled benign false positives from \textbf{30.26\%} to \textbf{22.05\%}, and external-judge and manual evaluations show substantially less harmful residual behavior.

  \section{Related Work}

  \subsection{Jailbreak Attacks on LLMs}

  Jailbreak attacks aim to bypass safety alignment and induce harmful or policy-violating outputs. Early attacks often relied on manually written role-play or prompt-injection templates~\cite{wei2023jailbroken}, while later attacks became increasingly automated through adversarial suffixes~\cite{zou2023universal}, search-based black-box optimization~\cite{chao2023pair}, stealthier optimization strategies such as AutoDAN~\cite{liu2023autodan}, and prompt decomposition methods such as DrAttack~\cite{li2024drattack}. Benchmark suites such as HarmBench broaden this evaluation space across diverse attack families~\cite{mazeika2024harmbench}, and multi-turn datasets such as SafeMTData highlight the importance of robustness beyond single-turn prompts~\cite{ren2024derail}. The present work is designed for this broader jailbreak setting rather than for a single prompt family.

  \subsection{Safety Defenses for LLMs}

  Current defenses include input filtering, output classification, safe decoding, refusal fine-tuning, and adversarial training. Guard-model systems such as Llama-Guard screen prompts or responses at the text level~\cite{inan2023llamaguard}, while decoding-time approaches such as SafeDecoding directly suppress harmful continuations~\cite{xu2024safedecoding}. Training-time approaches include instruction-following and refusal alignment~\cite{ouyang2022instructgpt}, latent adversarial training~\cite{sheshadri2024latent}, interpretable moderation pipelines~\cite{li2024safetyanalyst}, and reasoning-oriented defense strategies such as ARMOR~\cite{zhao2025armor}. These approaches can be effective under specific threat models, but their mechanism is usually external to the model's internal representations. NeuroArmor differs by explicitly testing whether the model's hidden state is consistent with safe variants and then intervening on that state so that generation moves either toward refusal or toward a helpful, non-harmful region when necessary.

  \subsection{Representation Engineering and Runtime Intervention}

  Representation engineering suggests that internal model states contain interpretable directions associated with concepts such as truthfulness, sentiment, or safety~\cite{zou2023representation}, and runtime steering methods exploit this idea by injecting such directions during inference~\cite{turner2023actadd,panickssery2023caa,arditi2024refusal}. NeuroArmor is related to this line of work, but it is not equivalent to it. Its central claim is that \textbf{inconsistency relative to prompt-specific safe variants provides a more defensible trigger for intervention than unconditional steering alone}.

  \subsection{Mechanistic Views of Alignment and Jailbreak}

  Recent work has increasingly analyzed alignment and jailbreak through hidden states, activation differences, and internal safety directions~\cite{zou2023representation,arditi2024refusal}. This line of work motivates our interpretation: in the evaluated models and datasets, successful jailbreaks are associated with measurable deviation from safety-conditioned internal representations. NeuroArmor therefore sits at the intersection of runtime defense and mechanism analysis.

  \section{Method}

  \begin{figure*}[t]
    \centering
    \includegraphics[width=\textwidth]{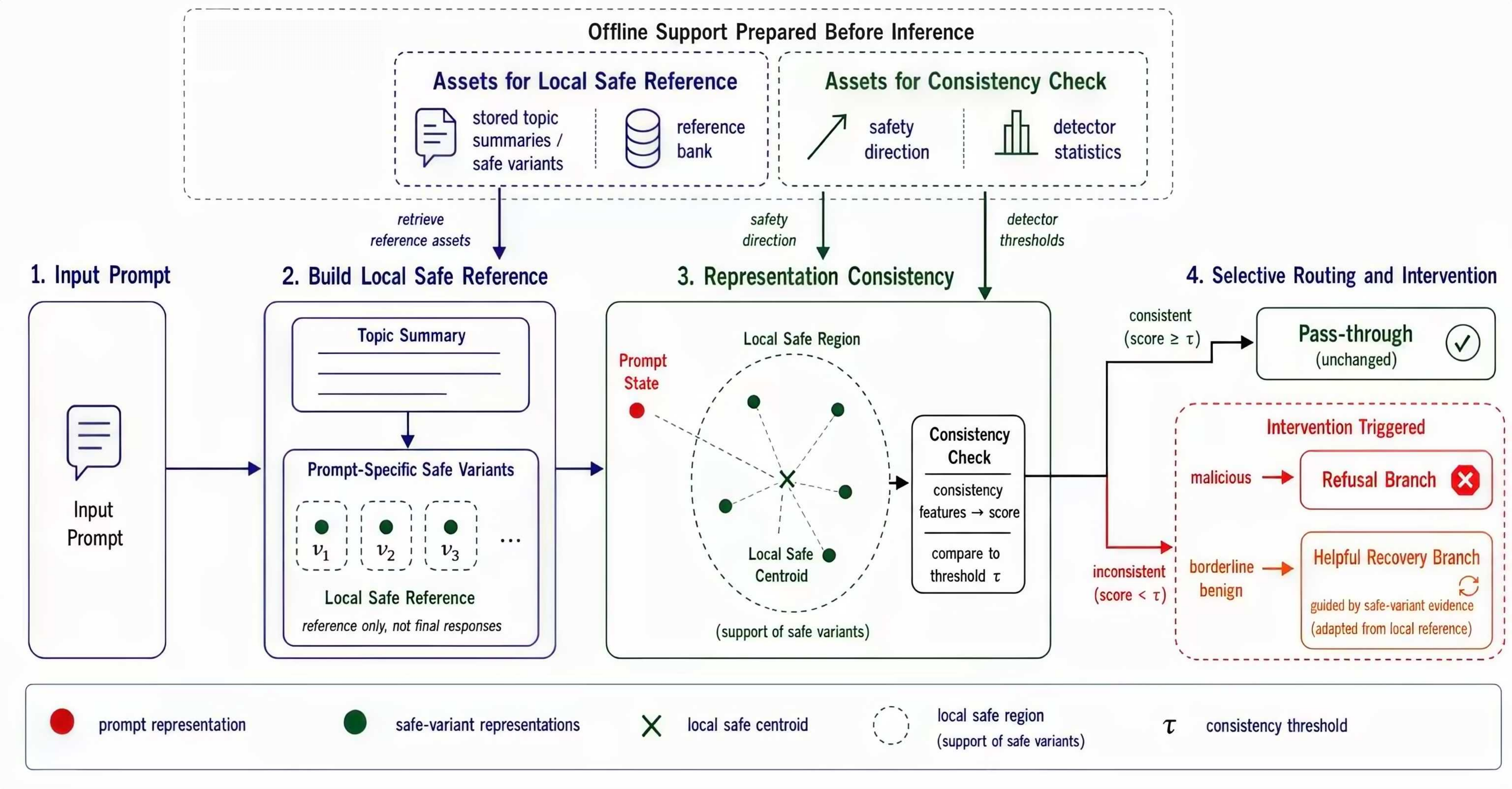}
    \caption{Overview of NeuroArmor at inference time. For each prompt, the system derives or retrieves a topic summary, builds or retrieves prompt-specific safe variants as a local safe reference, compares the prompt state against that reference in representation space, and intervenes only when inconsistency is detected. Consistent prompts pass through unchanged, while triggered cases are routed either to a refusal branch for malicious prompts or to a helpful recovery branch for borderline benign prompts. The small top inset summarizes offline-built support prepared before held-out inference, including stored topic summaries and safe variants, a reference bank, a safety direction, and detector statistics; at runtime, the system may retrieve these stored assets or fall back to deterministic local safe-variant construction, without using held-out labels or annotated harmful objectives.}
    \label{fig:overview}
  \end{figure*}

  \subsection{Method Overview}

  Given an input $x$, a model $f_{\theta}$ produces a response $y=f_{\theta}(x)$. We consider both malicious inputs and benign inputs, and the design goal is to suppress malicious outputs while preserving useful behavior on benign but sensitive requests. In all main experiments, the detector does not access ground-truth malicious/benign labels, attack families, or ground-truth intervention choices at inference time.

  At a high level, NeuroArmor runs three steps at inference time: it builds safe variants from the input prompt, uses a consistency detector to compare the prompt state with those variants, and, if needed, routes the prompt to either a refusal branch or a helpful recovery branch. In the reported implementation, the representation-consistency features provide the primary trigger signal, while fixed variant-construction rules and lightweight wrapper or multi-turn cues support a single runtime policy across prompt types. The core question is whether this combination of prompt-specific reference variants, consistency detection, and selective routing improves the malicious-suppression/benign-calibration trade-off beyond universal intervention baselines.

  \subsection{Safe Variants}

  For each input prompt, NeuroArmor constructs $K$ \textbf{safe variants}. These variants preserve the topic of the raw prompt while removing executable harmful assistance. In practice, they reframe the request as risk analysis, refusal explanation, or lawful alternative seeking. At inference time, the same high-level rule is applied to benign and malicious inputs. The system derives a short topic summary from raw text, reuses a matched bank entry when available, and otherwise falls back to a deterministic local rule. This keeps the method free of benchmark labels, attack-family metadata, and annotated harmful objectives at test time.

  The variants are not returned to the user as answers. Instead, they act as prompt-specific \textbf{internal reference points}. NeuroArmor compares the input state to those reference variants to determine whether the prompt remains inside a safe region and, if not, which safe destination is most appropriate. When no stored phase-1 entry is available, the same fixed fallback templates are used to instantiate the local safe reference from raw text before the consistency check is applied. Detailed construction rules are deferred to Appendix~\ref{app:variant-details}; a compact pipeline sketch is provided in Appendix~\ref{app:pipeline-sketch}.

  \subsection{Safety Direction Construction}

  Let $h_{\ell,T}(x)$ denote the residual-stream state at layer $\ell$ and the final prompt token position $T$. Given harmful-safe pairs $(x_a^i, x_b^i)$, we define the safety direction as

  \begin{equation}
  v_{\mathrm{safe}}^{(\ell)}
  =
  \mathrm{Norm}\left(
  \frac{1}{N}\sum_{i=1}^{N}
  \left[
  h_{\ell,T}(x_b^i)-h_{\ell,T}(x_a^i)
  \right]
  \right).
  \end{equation}

  This direction represents the average shift from harmful internal states toward safe-aligned internal states. The main experiments use a fixed target layer shared by the detector and intervention modules to keep comparisons fair against direct-steering baselines.

  \subsection{Consistency Detection}

  For a test input $x$, let $\mathcal{B}(x)=\{b_1,\ldots,b_K\}$ denote its safe-variant set. NeuroArmor computes three detector features and one auxiliary routing statistic:

  \begin{enumerate}
  \item the projection gap between the input and the mean variant projection,
  \item the cosine-consistency gap between the input and the variants,
  \item the dispersion of the joint state set $\{x,b_1,\ldots,b_K\}$,
  \item the signed projection of the input onto the safety direction.
  \end{enumerate}

  Formally, if

  \begin{equation}
  c(x)=\frac{h_{\ell,T}(x)^\top v_{\mathrm{safe}}}{\|v_{\mathrm{safe}}\|_2^2},
  \end{equation}

  then the projection gap is

  \begin{equation}
  \delta_c(x)
  =
  \left|
  c(x)-\frac{1}{K}\sum_{k=1}^{K}c(b_k)
  \right|,
  \end{equation}

  and the cosine-consistency gap is

  \begin{equation}
  \begin{split}
  \delta_{\cos}(x)
  &=
  \left|
  \cos(h_{\ell,T}(x),v_{\mathrm{safe}})
  \right. \\
  &\qquad \left.
  -\frac{1}{K}\sum_{k=1}^{K}\cos(h_{\ell,T}(b_k),v_{\mathrm{safe}})
  \right|.
  \end{split}
  \end{equation}

  The learned detector uses the three-feature vector $z(x)=[\delta_c(x),\delta_{\cos}(x),\mathrm{variance}(x)]$. The signed projection $c(x)$ is retained only as an auxiliary routing statistic. In the reported implementation, a One-Class SVM~\cite{scholkopf2001estimating} is trained on stored reference features exported as \texttt{svm\_train\_features.npy}; full training details are deferred to Appendix~\ref{app:detector-reanchoring-details}. At test time, its output is combined with percentile-based outlier rules and lightweight multi-turn and wrapper heuristics. We use this lightweight detector to keep the focus on the consistency signal and routing policy rather than detector complexity.

  \subsection{Routing and Intervention}

  When the detector flags an anomaly, NeuroArmor routes the prompt to one of two branches. The \textbf{refusal branch} handles malicious anomalies. It applies an intervention that moves the hidden state back toward the safe direction or safe centroid so that generation turns into refusal. The \textbf{helpful recovery branch} handles borderline benign prompts, including prompts whose surface form may still look suspicious. Instead of refusing by default, it uses safe-variant evidence to recover a useful, non-harmful interpretation.

  In the reported implementation, the intervention is applied at a fixed residual-stream layer shared by detection and intervention. The hook edits the hidden state at the last prompt position during generation. Multi-turn cases use a safe continuation rewrite for the next assistant turn. Single-turn cases are routed between the refusal branch and the helpful recovery branch by the same detector outputs and safe-variant evidence. Exact thresholds, branch-selection rules, and coefficient schedules are deferred to Appendix~\ref{app:detector-reanchoring-details}.

  This target-layer choice is an operating point rather than a universal claim about where intervention must occur. In the reported Llama-3 sensitivity sweep, earlier layers reduce benign blocking sharply but leave materially higher ASR, whereas the final block gives the strongest malicious suppression at the cost of harsher benign calibration (Appendix Table~\ref{tab:layer-sensitivity}).

  Our claim is not that every low-level intervention operator is individually necessary, but that using safe variants to choose between the refusal branch and the helpful recovery branch works better than unconditional steering or weak heuristic patching, especially when the goal is to suppress malicious outputs without inflating benign refusal.

  \section{Experiments}

  \subsection{Experimental Setup }

  At inference time, the system builds safe variants, measures representation consistency, and intervenes only when the detector fires. The headline experiments use Llama-3-8B-Instruct with white-box activation access, and include matched Gemma-2-9B results under the same setup as supporting cross-model evidence~\cite{llama32024herd,gemma22024}. The main held-out malicious pool contains 2416 held-out jailbreak examples from the main malicious test file, spanning families such as AutoDAN, PAIR, DrAttack, GCG, Base64, SAA, IJP, Zulu, and Puzzler. The shared benign pool contains 2925 examples from XS-Test and OR-Bench-Hard~\cite{cui2024orbench,rottger2023xstest}.

  We refer to the full intervention system as \texttt{NeuroArmor-ReAnchor} and to the detector-only logging run as \texttt{NeuroArmor-Detect}. We report malicious attack success rate (ASR), the fraction of malicious prompts whose final response is not blocked, and benign false positive rate (FPR), the fraction of benign prompts whose final response is blocked. We complement these with matched held-out comparisons, GPT-4 external-judge evaluations, and manual harmfulness review. We compare against Vanilla, Direct Safety Steering, Runtime Intervention Baseline, and Llama-Guard~\cite{inan2023llamaguard}. Unless otherwise noted, phase-2 runs use model-specific phase-1 reference sets with $K=3$ safe variants, a fixed final-block hook layer, and a One-Class SVM over the consistency features. The reported thresholds, routing rules, and intervention coefficients are fixed before held-out test evaluation. Full split definitions and exact routing thresholds are deferred to Appendix~\ref{app:variant-details}, Appendix~\ref{app:data-splits}, and Appendix~\ref{app:detector-reanchoring-details}.

  \subsection{Main Held-Out Results}

  Table~\ref{tab:matched-heldout-summary} provides the aggregate Llama-3 held-out comparison, including the matched Llama-Guard-3 reference. Table~\ref{tab:main-textkey} then breaks the main held-out malicious evaluation down by attack family rather than collapsing the entire pool into a single scalar. For the main Llama-3-8B-Instruct comparison, the direct-steering and runtime-intervention rows are matched back to the exact malicious rows used by the NeuroArmor evaluation.

  \begin{table}[t]
\centering
\small
\setlength{\tabcolsep}{4pt}
\begin{tabular}{@{}lcc@{}}
\toprule
System & ASR (\%) $\downarrow$ & FPR (\%) $\downarrow$ \\
\midrule
Vanilla & 41.56 & 30.26 \\
NeuroArmor-ReAnchor & \textbf{1.57} & \textbf{22.05} \\
Direct Safety Steering & 23.92 & 33.54 \\
Runtime Intervention Baseline & 24.92 & 33.40 \\
Llama-Guard-3 & 26.61 & \textbf{9.61} \\
\bottomrule
\end{tabular}
\caption{Aggregate Llama-3 comparison on the matched held-out malicious pool and the corresponding benign evaluation runs. Lower is better.}
\label{tab:matched-heldout-summary}
\end{table}

  \begin{table*}[t]
\centering
\small
\setlength{\tabcolsep}{2.4pt}
\renewcommand{\arraystretch}{1.06}
\resizebox{\textwidth}{!}{%
\begin{tabular}{@{}llccccccccc@{}}
\toprule
Model & Defense & \multicolumn{9}{c}{Jailbreak Attack Families $\downarrow$} \\
& & GCG & AutoDAN & DrAttack & PAIR & SAA & IJP & Base64 & Zulu & Puzzler \\
\midrule
Llama-3-8B-Instruct & Vanilla & 5.67 (17/300) & 3.00 (9/300) & 22.33 (67/300) & 17.33 (52/300) & 21.33 (64/300) & 66.00 (198/300) & 100.00 (300/300) & 93.67 (281/300) & 100.00 (16/16) \\
 & NeuroArmor-Detect & 5.67 (17/300) & 3.00 (9/300) & 22.33 (67/300) & 17.33 (52/300) & 21.33 (64/300) & 66.00 (198/300) & 100.00 (300/300) & 93.67 (281/300) & 100.00 (16/16) \\
 & NeuroArmor-ReAnchor & \textbf{0.00 (0/300)} & \textbf{0.00 (0/300)} & \textbf{0.00 (0/300)} & \textbf{3.67 (11/300)} & \textbf{0.00 (0/300)} & \textbf{8.67 (26/300)} & \textbf{0.00 (0/300)} & \textbf{0.33 (1/300)} & \textbf{0.00 (0/16)} \\
 & Direct Safety Steering & 7.33 (22/300) & 4.67 (14/300) & 20.67 (62/300) & 19.00 (57/300) & 20.33 (61/300) & 61.33 (184/300) & 0.33 (1/300) & 54.00 (162/300) & 93.75 (15/16) \\
 & Runtime Intervention Baseline & 5.67 (17/300) & 5.00 (15/300) & 22.33 (67/300) & 17.00 (51/300) & 20.00 (60/300) & 62.00 (186/300) & 1.67 (5/300) & 62.67 (188/300) & 81.25 (13/16) \\
\midrule
Gemma-2-9B & Vanilla & 2.67 (8/300) & 13.33 (40/300) & 23.00 (69/300) & 26.00 (78/300) & 99.33 (298/300) & 88.33 (265/300) & 75.00 (225/300) & 98.00 (294/300) & 100.00 (16/16) \\
 & NeuroArmor-Detect & 2.67 (8/300) & 13.33 (40/300) & 23.00 (69/300) & 26.00 (78/300) & 99.33 (298/300) & 88.33 (265/300) & 75.00 (225/300) & 98.00 (294/300) & 100.00 (16/16) \\
 & NeuroArmor-ReAnchor & \textbf{0.00 (0/300)} & \textbf{0.00 (0/300)} & \textbf{0.33 (1/300)} & \textbf{3.67 (11/300)} & \textbf{0.00 (0/300)} & \textbf{11.33 (34/300)} & \textbf{0.00 (0/300)} & \textbf{0.00 (0/300)} & \textbf{0.00 (0/16)} \\
\bottomrule
\end{tabular}
}
\caption{Main ASR comparison broken down by jailbreak attack family. Each entry reports attack success rate as a percentage, followed by the exact successful-jailbreak count in parentheses. Lower is better. The Llama-3 block includes matched direct-steering and runtime-intervention baselines; the Gemma-2 block provides cross-model support under the same evaluation setting.}
\label{tab:main-textkey}
\end{table*}

  On Llama-3-8B-Instruct, NeuroArmor-ReAnchor reduces ASR from \textbf{41.56\%} to \textbf{1.57\%}, while pooled benign FPR on the shared benign pool improves from \textbf{30.26\%} to \textbf{22.05\%}. The matched aggregate comparison in Table~\ref{tab:matched-heldout-summary} makes the trade-off explicit: Direct Safety Steering reaches \textbf{23.92\%} ASR with \textbf{33.54\%} FPR, Runtime Intervention Baseline reaches \textbf{24.92\%} with \textbf{33.40\%} FPR, and Llama-Guard-3 reaches \textbf{26.61\%} ASR with \textbf{9.61\%} FPR. The Detect row is expected to match Vanilla on ASR because it logs detector behavior without changing generation.

  The family-level breakdown shows the same qualitative pattern across attack families. For Llama-3-8B-Instruct, the hardest vanilla families are IJP, Base64, Zulu, and Puzzler, whereas NeuroArmor-ReAnchor suppresses nearly all families to zero or near-zero, leaving only small residual exposure on PAIR and IJP. The Gemma-2-9B block shows the same qualitative pattern and provides supporting cross-model evidence. Supplementary borderline-benign and adaptive-attack results are reported in Appendix~\ref{app:sensitivity-robustness}.

  \begin{table*}[t]
\centering
\scriptsize
\setlength{\tabcolsep}{3.5pt}
\resizebox{\textwidth}{!}{%
\begin{tabular}{llcccccc}
\toprule
Model & Defense & \multicolumn{3}{c}{Small Malicious Sets (ASR $\downarrow$)} & \multicolumn{3}{c}{Benign Sets (FPR $\downarrow$)} \\
& & AdvBench & HarmfulBehaviors & SafeMTData & XS-1356 & XS-safe & OR-Bench-Hard \\
\midrule
Llama-3-8B-Instruct & Vanilla & 1.54 (8/520) & 7.00 (7/100) & 83.67 (502/600) & 1.11 (15/1356) & 3.20 (8/250) & 65.35 (862/1319) \\
 & NeuroArmor-Detect & 1.54 (8/520) & 7.00 (7/100) & 83.67 (502/600) & 1.11 (15/1356) & 3.20 (8/250) & 65.35 (862/1319) \\
 & NeuroArmor-ReAnchor & \textbf{0.00 (0/520)} & \textbf{0.00 (0/100)} & \textbf{13.83 (83/600)} & \textbf{0.37 (5/1356)} & \textbf{4.00 (10/250)} & \textbf{47.76 (630/1319)} \\
 & Direct Safety Steering & 1.54 (8/520) & 8.00 (8/100) & 80.33 (482/600) & 4.50 (61/1356) & 5.60 (14/250) & 68.69 (906/1319) \\
 & Runtime Intervention Baseline & 1.54 (8/520) & 7.00 (7/100) & 79.50 (477/600) & 4.20 (57/1356) & 6.40 (16/250) & 68.54 (904/1319) \\
Gemma-2-9B & Vanilla & 0.96 (5/520) & 3.00 (3/100) & 87.67 (526/600) & 7.37 (100/1356) & 15.60 (39/250) & 81.27 (1072/1319) \\
 & NeuroArmor-Detect & 0.96 (5/520) & 3.00 (3/100) & 87.67 (526/600) & 7.37 (100/1356) & 15.60 (39/250) & 81.27 (1072/1319) \\
 & NeuroArmor-ReAnchor & \textbf{0.00 (0/520)} & \textbf{0.00 (0/100)} & \textbf{26.17 (157/600)} & \textbf{5.90 (80/1356)} & \textbf{13.20 (33/250)} & \textbf{50.19 (662/1319)} \\
\bottomrule
\end{tabular}
}
\caption{Single-dataset results outside the large jailbreak-family block. The left block reports ASR on smaller malicious benchmarks, while the right block reports FPR on benign slices. Each entry is shown as percentage with exact count fraction in parentheses. Lower is better throughout.}
\label{tab:dataset-slices-textkey}
\end{table*}

  Table~\ref{tab:dataset-slices-textkey} clarifies where the benign-side error is concentrated. On the malicious side, NeuroArmor-ReAnchor drives AdvBench and HarmfulBehaviors to \textbf{0.00\%} ASR for both Llama-3-8B-Instruct and Gemma-2-9B, while SafeMTData remains the hardest standalone slice at \textbf{13.83\%} and \textbf{26.17\%}, respectively~\cite{ren2024derail}. On the benign side, the remaining false-positive burden is concentrated on OR-Bench-Hard, whereas the two XS-Test slices remain materially lower. This pattern is consistent with a concentration of benign errors on borderline benign prompts rather than a broad collapse on ordinary benign instructions.

  These slice results also show that the reported gains are not limited to exact phase-1 bank matches. The bank is fixed before testing, yet the held-out multi-turn attack slice and the borderline-benign slice improve only when detection and intervention are used together. Appendix Table~\ref{tab:overlap-audit} reports the corresponding exact-overlap audit and held-out exact-match rate to stored phase-1 variants.

  \subsection{Behavioral Evaluation}
  \label{subsec:behavioral-results}

  To complement ASR, we apply a fixed GPT-4 external judge~\cite{openai2023gpt4} with a harmfulness rating scheme. The judge receives the original malicious prompt and the final model response, then returns one label from \{\texttt{operational\_harmful}, \texttt{refusal}, \texttt{safe\_answer}, \texttt{non\_harmful\_offtarget}, \texttt{degenerate}, \texttt{answer}\}. We define \textbf{Harm N} as the number of malicious examples labeled \texttt{operational\_harmful} and \textbf{Judge ASR} as \textbf{Harm N} divided by the number of evaluated malicious examples.

  Table~\ref{tab:external-judge} reports the external-judge evaluation on the same unified malicious pool of \texttt{2416} held-out examples. For \texttt{Llama-3-8B}, harmful rows drop from \textbf{620/2416} under \texttt{Vanilla} to \textbf{21/2416} under \texttt{NeuroArmor-ReAnchor}; for Gemma, they drop from \textbf{743/2416} to \textbf{32/2416}. This behavioral view is aligned with the ASR result: selective intervention not only blocks more malicious generations, but also shifts the remaining non-blocked outputs away from operationally harmful behavior. The fuller manual category breakdown is reported in Appendix~\ref{app:behavioral-rubric}.

  \begin{table}[t]
\centering
\scriptsize
\setlength{\tabcolsep}{2.6pt}
\resizebox{\columnwidth}{!}{%
\begin{tabular}{@{}llcrrrr@{}}
\toprule
Model & System & N & ASR & Harm N & Judge ASR & Red. (\%) \\
\midrule
\texttt{Llama-3-8B} & Vanilla & 2416 & 41.56 & 620 & 25.66 & 0.00 \\
\texttt{Llama-3-8B} & NeuroArmor-Detect & 2416 & 41.56 & 620 & 25.66 & 0.00 \\
\texttt{Llama-3-8B} & NeuroArmor-ReAnchor & 2416 & \textbf{1.57} & 21 & 0.87 & 96.61 \\
\texttt{Llama-3-8B} & Direct Safety Steering & 2416 & 23.92 & 443 & 18.34 & 28.55 \\
\texttt{Llama-3-8B} & Runtime Intervention Baseline & 2416 & 24.92 & 530 & 21.94 & 14.52 \\
\texttt{Llama-3-8B} & Llama-Guard-3 & 2416 & 26.61 & 410 & 16.97 & 33.87 \\
\midrule
\texttt{Gemma-2-9B} & Vanilla & 2416 & 53.52 & 743 & 30.75 & 0.00 \\
\texttt{Gemma-2-9B} & NeuroArmor-Detect & 2416 & 53.52 & 743 & 30.75 & 0.00 \\
\texttt{Gemma-2-9B} & NeuroArmor-ReAnchor & 2416 & \textbf{1.90} & 32 & 1.32 & 95.69 \\
\bottomrule
\end{tabular}
\par}
\caption{GPT-4 external-judge results on the unified main held-out malicious pool of \texttt{2416} examples. The \texttt{ASR} column is reported on the same pool.}
\label{tab:external-judge}
\end{table}

  The same ranking also holds in the full aligned comparison deferred to Appendix Table~\ref{tab:behavioral-alignment}, where \texttt{NeuroArmor-ReAnchor} remains the strongest setting under all three views.

  \subsection{Ablation Results}

  Table~\ref{tab:ablation-textkey} summarizes the held-out ablation on the same main held-out evaluation setting used in the main results.

  \begin{table}[t]
\centering
\small
\begin{tabular}{lcc}
\toprule
Configuration & ASR (\%) & FPR (\%) \\
\midrule
Full NeuroArmor-ReAnchor & 1.57 & 22.05 \\
Detector only & 39.43 & 33.44 \\
Re-anchor only & 0.00 & 100.00 \\
No-variant detector & 30.32 & 34.33 \\
Projection-only detector & 6.02 & 50.76 \\
Direction-only re-anchor & 23.48 & 35.29 \\
\bottomrule
\end{tabular}
\caption{Held-out ablation on the same main held-out evaluation setting used in the main results. The table tests selective triggering, safe-variant conditioning, and full re-anchoring beyond a direct safety-direction push.}
\label{tab:ablation-textkey}
\end{table}

  This main held-out ablation supports four claims. First, \texttt{Detector only} leaves ASR high at \textbf{39.43\%}, so anomaly logging without intervention is insufficient. Second, \texttt{Re-anchor only} reaches \textbf{0.00\%} ASR only by collapsing benign usability to \textbf{100.00\% FPR}, which makes clear why trigger/action separation matters: without routing, all flagged cases are pushed toward the same refusal-like endpoint. Third, \texttt{Direction-only re-anchor} remains much weaker than the full system, at \textbf{23.48\%} ASR and \textbf{35.29\%} FPR versus \textbf{1.57\%} and \textbf{22.05\%}, so the gain is not explained by a uniform safety-direction push alone. Fourth, the detector side also contributes materially to the trade-off. Removing safe variants degrades the full row to \textbf{30.32/34.33}, while a projection-only trigger reaches \textbf{6.02/50.76}, indicating that prompt-specific safe references and multi-feature consistency checks both matter for calibration. Taken together, Table~\ref{tab:ablation-textkey} supports selective triggering, variant-conditioned routing, and trigger/action separation.

  Appendix Table~\ref{tab:operator-ablation} provides a separate operator-level ablation with the same qualitative pattern: weakening the intervention stack degrades the reported trade-off, which is more consistent with a coordinated detector-and-routing design than with any single low-level operator acting alone.

  \subsection{Utility Preservation}

  Appendix Table~\ref{tab:utility} reports the full utility breakdown. No regression is observed on the evaluated academic benchmarks. Because the detector does not trigger on these tasks, we read this result conservatively as absence of observed side effects rather than as evidence of broad utility gain, and treat the borderline-benign and dataset-slice analyses above as the more informative utility-side evidence.

  \FloatBarrier
  \section{Analysis}

  \subsection{Steering vs. Routing}

  The direct-steering baseline tests the simplest alternative explanation: perhaps any safety vector helps. The results argue against that explanation. \texttt{Direct Safety Steering} improves over \texttt{Vanilla}, but the improvement is substantially smaller than that of \texttt{NeuroArmor-ReAnchor}. The ablation results sharpen the same point from a second angle: removing safe variants materially hurts ASR, and direction-only intervention remains far weaker than the full system. Under the reported fixed runtime policy, these comparisons are more consistent with prompt-specific safe references driving the trigger-and-target decision than with a generic safety push alone.

  This interpretation is also consistent with the Appendix $K$-sweep. Moving from \texttt{K=1} to \texttt{K=3} yields a large ASR reduction, while \texttt{K=4} and \texttt{K=5} bring only marginal additional movement (Appendix Table~\ref{tab:k-sensitivity}). The method therefore appears to benefit from multiple local reference variants, but is already near saturation at the reported \texttt{K=3} operating point.

  \subsection{Recovery on Borderline Benign Prompts}

  The benign-side behavior is also consistent with this routing view. Once safe variants serve not only as anomaly references but also as helpful, non-harmful destinations, ambiguous sensitive requests whose surface form looks suspicious or dual-use no longer need to collapse into refusal by default. Instead, the model can be directed toward a helpful interpretation of the same topic. Within the reported policy, this provides a natural account of why benign calibration can improve even though intervention is still applied selectively at runtime. The pattern is consistent with a two-branch effect: the refusal branch blocks malicious cases, while the helpful recovery branch recovers borderline benign prompts.

  The slice-level results in Table~\ref{tab:dataset-slices-textkey} support this interpretation. The remaining false-positive burden is concentrated on \texttt{OR-Bench-Hard}, whereas the two \texttt{XS-Test} slices remain materially lower. This pattern suggests that the benign FPR is driven primarily by difficult boundary cases rather than by a broad collapse on ordinary benign instructions.

  \FloatBarrier
  \subsection{Mechanistic Evidence}

  The mechanism claim in this paper is supported by more than the final ASR/FPR trade-off alone. In the evaluated setting, the same prompt-specific safe reference that enters the runtime policy also yields measurable representation-level separation before intervention and directed movement back toward the safe region after intervention. This interpretation is consistent with prior evidence that safety-relevant structure can be read from and steered within hidden states~\cite{zou2023representation,arditi2024refusal}.

  \begin{figure}[H]
    \centering
    \includegraphics[width=\columnwidth]{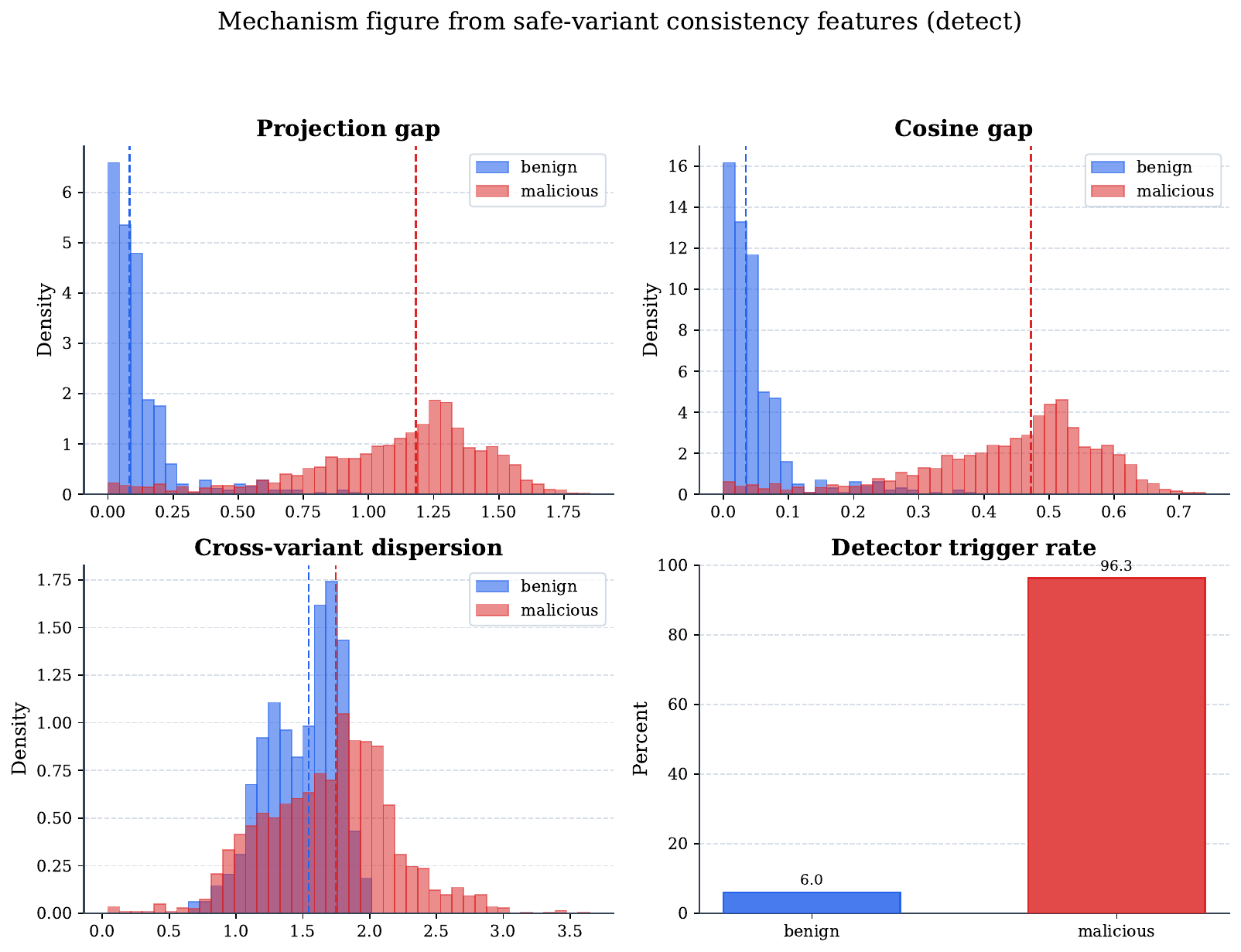}
    \caption{Held-out consistency-feature distributions for benign and malicious prompts. The figure visualizes the detector-feature separation used by the runtime policy.}
    \label{fig:consistency}
  \end{figure}

  Figure~\ref{fig:consistency} provides a representation-level view of the same detector features used by the reported runtime policy. The separation between benign and malicious rows is consistent with using representation inconsistency as a trigger signal in the reported setting. The visualization is generated from phase-2 feature rows for the reported evaluation system, and the plotted histograms show empirical density over the held-out evaluation rows rather than normalized latent dimensions chosen after inspection.

  \begin{figure}[t]
    \centering
    \includegraphics[width=\columnwidth]{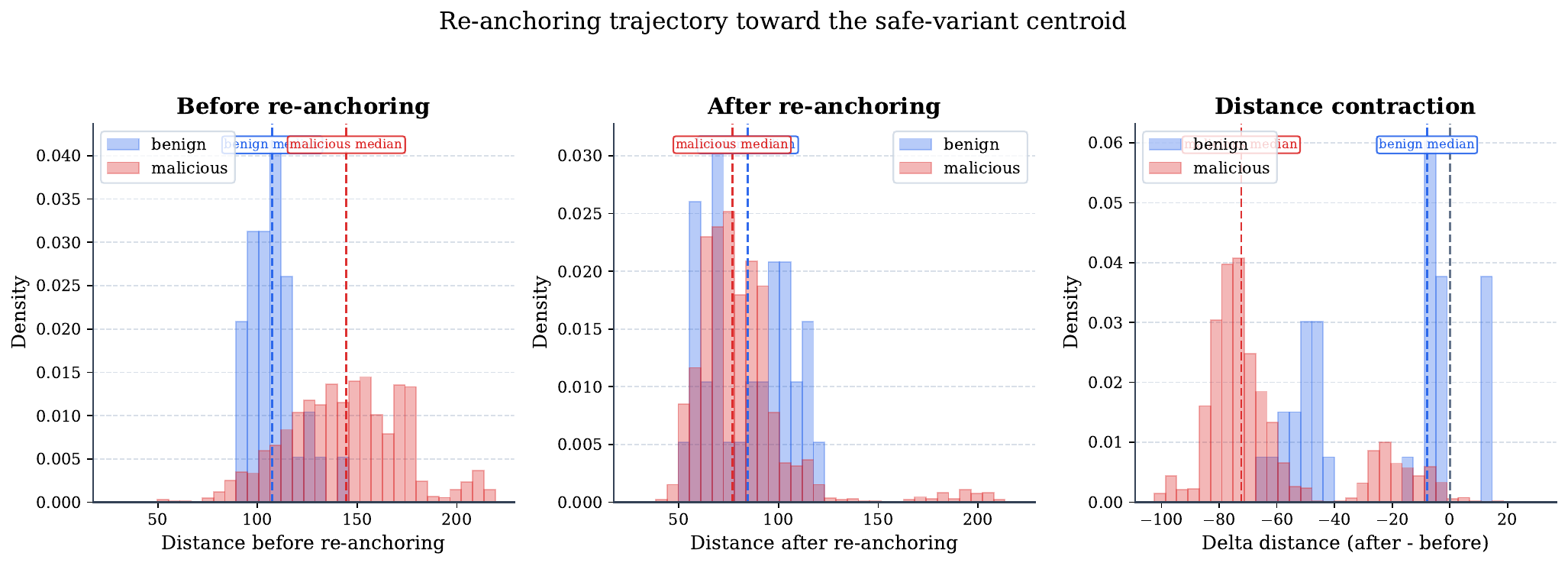}
    \caption{Held-out intervention trajectories toward the safe-variant centroid. The figure visualizes pre/post movement toward the safe reference region under the reported intervention pipeline.}
    \label{fig:reanchoring-trajectory}
  \end{figure}

  Figure~\ref{fig:reanchoring-trajectory} reports pre/post distances to the safe-variant centroid over triggered held-out cases under the same phase-2 pipeline. This pattern is consistent with the interpretation that intervention redirects triggered states toward the safe reference region used by the policy. Together with Figure~\ref{fig:consistency}, it connects the consistency-based trigger to the intervention target used by the reported system.

  The available evidence supports the main claim, but it does not justify stronger claims such as universal necessity of detector gating or universal necessity of alignment-based adjustment. More conservatively, the present experiments support representation inconsistency as a useful empirical trigger within the reported runtime policy in the evaluated models and datasets, not as a universal or exclusive causal explanation of jailbreak success. At the same time, finer-grained separation from template design, threshold tuning, refusal surface form, and other evaluation-side choices remains an open target for follow-up study, even though the matched baselines, overlap checks, and full-pool manual review reduce those confounds substantially. Overall, the results support NeuroArmor as a routing-based defense grounded in prompt-specific safe references and consistency-based intervention.

  \section{Conclusion}

  NeuroArmor supports a narrower but more concrete view of jailbreak defense: prompts that successfully bypass alignment often do so by departing from the safety-conditioned internal representations induced by safe variants. Measuring that departure provides a runtime signal for deciding when intervention is warranted, while the variants themselves provide structured safe targets for that intervention.

  The reported results indicate that this routing-based setup is materially stronger than unconditional steering or weak heuristic patching in the same setting. The most important implication is not only the reduction in malicious attack success, but also the accompanying behavioral evidence that the residual non-blocked responses shift away from operationally harmful categories. At the same time, the paper's own borderline-benign and adaptive-attack analyses make the remaining limitations clear: benign calibration is still difficult, and robust deployment claims would require broader evaluation beyond the present pools. Taken together, the evidence positions prompt-specific routing and intervention as a useful runtime defense strategy and as a concrete way to study how jailbreaks disrupt aligned internal representations.

  \section*{Limitations}

  NeuroArmor still has several clear limitations. First, it depends on white-box activation access and is therefore best suited to open-weight or otherwise white-box deployment settings. Second, the detector is based on a simple One-Class SVM for interpretability and practicality; stronger detectors may improve the trade-off further. Third, the ablations show that removing safe variants and weakening consistency-based triggering materially degrades the reported trade-off, but the present study does not yet fully disentangle that contribution from template design, threshold tuning, refusal surface form, or other evaluation-side choices. We reduce those confounds through exact implementation disclosure, matched per-example baselines, split-overlap checks, and full-pool manual category review, but we do not eliminate them. Fourth, the adopted split strategy enforces strict row-level train/validation/test separation, and the appendix reports zero exact-row overlap under that setting. Future work could extend this safeguard with stronger semantic deduplication guarantees across diverse upstream benchmarks. Fifth, adaptive attackers and richer borderline-benign tasks, especially multi-turn benign conversations and dual-use requests, still require dedicated evaluation; our own adaptive probe table already shows that detector-aware attacks can recover nontrivial ASR.

  \section*{Ethics Statement}

  This work studies jailbreak defense and uses harmful prompts only for safety evaluation. It does not release executable harmful instructions and avoids exposing operationally dangerous details. The goal of NeuroArmor is to reduce harmful generation, clarify how safety failures manifest internally, and improve the interpretability of runtime safety interventions. All datasets and models used in this paper are public research artifacts evaluated only in a research setting for safety analysis. The benchmark data may contain harmful or offensive content by design, but we do not collect new personal data and we avoid reproducing dangerous content beyond what is necessary to describe the evaluation setting.

  \FloatBarrier
  \bibliography{neuroarmor_acl_submission_refs}

  \appendix
  \sloppy

  \section{Safe-Variant Construction Details}
  \label{app:variant-details}

  The reported system uses \texttt{K=3} safe variants per prompt. The key intermediate object is a short topic summary $s$ that captures the request at a safe level of abstraction. The variants themselves are not open-ended model answers. Instead, NeuroArmor inserts $s$ into a small fixed template family. In the reported single-turn runs, these templates correspond to risk analysis, refusal explanation, and helpful or lawful-alternative framing. For multi-turn inputs, the system preserves the earlier turns and rewrites only the next assistant turn into a prevention-oriented safe continuation.

  The construction has two distinct stages that should not be conflated. \textbf{Phase 1} builds the offline reference bank and detector inputs from the jailbreak training pool only. \textbf{Phase 2} applies the resulting bank at held-out inference time. The main evaluation therefore does not require access to held-out benchmark labels, attack-family metadata, or annotated harmful-objective text at runtime.

  During \textbf{offline reference-bank construction}, the builder first deduplicates the jailbreak training pool and then assigns each retained row a short topic summary. The other malicious benchmark files in this project are not used to build the phase-1 bank; they are reserved for held-out evaluation slices. When a dataset-provided reference goal is available in the jailbreak training data, the builder may use that text directly after normalization, or use a constrained model-assisted summarizer to compress the attack text into a short neutral topic description. This summarizer is deterministic (\texttt{temperature=0.0}, \texttt{max\_new\_tokens=64}, tokenizer truncation at \texttt{max\_length=1024}) and is instructed to output only a brief description of the underlying objective, without instructions, refusal language, or policy explanation. A candidate summary is accepted only if it is short, non-empty, free of refusal boilerplate, and has lexical overlap with the available training reference text; otherwise the builder falls back to a normalized reference-goal or source-prompt string. In the Llama-3 phase-1 bank used here, \texttt{1096/1725} topic summaries (\texttt{63.5\%}) are accepted from the constrained summarizer and \texttt{629/1725} (\texttt{36.5\%}) use the deterministic fallback; the corresponding Gemma rates are \texttt{53.9\%} and \texttt{46.1\%}.

  During \textbf{held-out inference}, the system does not query annotated harmful objectives. It first normalizes the raw prompt and looks for a match in the stored phase-1 bank. If a match is found, it reuses the stored topic summary and its associated safe variants. If no stored match is found, the reported evaluation switches to a deterministic local fallback. Concretely, the raw prompt text is whitespace-normalized, truncated to \texttt{220} characters, and then passed through a fixed lexical screen. If the normalized text contains high-risk operational markers such as \texttt{hack}, \texttt{malware}, \texttt{phishing}, \texttt{bomb}, \texttt{weapon}, \texttt{bypass}, \texttt{steal}, \texttt{payload}, \texttt{base64}, or \texttt{ignore previous}, the system instantiates a fixed risk-facing template family: safety analysis, refusal explanation, and defensive review of the request. Otherwise it instantiates a fixed helpful template family: benign paraphrase, concise direct answer, and ambiguity-clarification prompts. This fallback is the reproducibility-critical held-out path. It is the path that executes whenever a test prompt misses the stored phase-1 bank. An optional online variant cache exists in code, but it is not required for the reported results. The same high-level rule is used for benign and malicious inputs: derive a short topic-level abstraction from raw text, construct fixed safe variants from it, and compare the resulting hidden states to the prompt state.

  The prompt-local selector is a literal keyword test over the raw lowercased prompt, and the risky-variant screen fires only when a structural cue and an operational harmful cue co-occur. Under this constrained pipeline, the legacy counters \texttt{json\_parse\_failure} and \texttt{semantic\_drift\_rejected} remain zero in the phase-1 reports. For the Llama-3 reference bank, the corresponding variant-effectiveness summary reports \texttt{5175} total variants, a \texttt{94.6\%} unique-variant ratio, a \texttt{0.17\%} risky-variant ratio under the internal heuristic screen, and average keyword overlap \texttt{0.59} between the stored training reference text and the accepted topic summary.

  \section{Data Splits and Evaluation Scope}
  \label{app:data-splits}

  The main held-out claim of the paper is anchored to the \texttt{jailbreak\_test\_test.jsonl} malicious pool and the shared benign pool used in the pooled benign FPR calculation. The split manifest fixes a \texttt{70/10/20} train/validation/test split, but the reported benign-side evaluation uses the full shared \texttt{2925}-example benign evaluation pool rather than the smaller per-file test slices. Phase-1 reference construction uses the jailbreak training split only, while the headline phase-2 tables read from the held-out validation/test files generated by the split manifest. The safe-variant reference bank itself is built only from \texttt{jailbreak\_test\_train.jsonl} (\texttt{8456}) after normalization and deduplication. The other malicious benchmark files in this project are not used to build phase-1 safe variants; they are reserved as held-out evaluation slices. In particular, \texttt{advbench\_train.csv}, \texttt{harmful-behaviors\_train.csv}, and \texttt{SafeMTData\_train.json} are evaluated in full in the malicious test suite rather than folded into the reference bank. The phase-1 builder then deduplicates the jailbreak training rows by normalized goal-or-jailbreak text before constructing safe variants, leaving \texttt{1725} unique malicious reference topics in the bank. Benign data participate separately: phase-1 uses a deduplicated benign calibration source distinct from the \texttt{2925}-example evaluation pool, and the detector training file includes only a small \texttt{40}-row benign calibration supplement in addition to the safe-side reference features. At inference time, held-out prompts may either match a stored phase-1 entry or use deterministic templates, but the evaluation pipeline does not consult held-out benchmark labels or annotated harmful-objective metadata when deciding whether to intervene. Appendix Table~\ref{tab:overlap-audit} reports zero exact-row overlap across split boundaries and an \texttt{18.00\%} held-out exact-match rate to stored phase-1 variants, so deterministic construction remains the majority path at inference time.

  For the matched Llama-3 baselines, we additionally restrict the full baseline detail files to the exact held-out rows used by the main claim. The matching key is the tuple \texttt{(label, source\_dataset, text, goal)}, and the matching procedure is documented in \texttt{compute\_matched\_baseline\_slices.py}. We do not claim a stronger proof of zero semantic overlap across all upstream benchmarks beyond this row-level split and deduplication procedure; Appendix Table~\ref{tab:overlap-audit} reports the corresponding overlap check under the same split setting.

  \section{Detector and Intervention Details}
  \label{app:detector-reanchoring-details}

  The detector is intentionally simple. It uses a \texttt{StandardScaler} followed by a One-Class SVM with RBF kernel, \texttt{nu=0.01}, and \texttt{gamma=0.5}, operating on three representation-consistency features: projection gap \texttt{c\_delta}, cosine-consistency gap \texttt{cos\_delta}, and cross-state dispersion \texttt{variance}. The signed projection $c$ is not part of the learned feature vector, but is retained as an auxiliary routing statistic.

  For compact notation in the remainder of this appendix subsection, we use the following aliases:
  \begin{center}
  \small
  \begin{tabular}{@{}ll@{}}
  $s_d$ & \texttt{detector\_score} \\
  $s_w$ & \texttt{wrapper\_risk\_score} \\
  $s_m$ & \texttt{multiturn\_risk\_score} \\
  $q_c$ & $\delta_{c,\mathrm{p95}}$ \\
  $m_c$ & $\delta_{c,\mathrm{p50}}$ \\
  $q_{\cos}$ & $\delta_{\cos,\mathrm{p95}}$ \\
  $q_v$ & $\mathrm{variance}_{\mathrm{p95}}$ \\
  \end{tabular}
  \end{center}

  Training uses the stored phase-1 file \texttt{svm\_train\_features.npy}. For the Llama-3 reference set used in these experiments, this file has shape \texttt{(5215, 3)}: \texttt{5175} rows come from safe-side within-set reference statistics over the \texttt{1725} malicious reference topics and their \texttt{K=3} safe variants, and \texttt{40} rows come from a small benign calibration supplement drawn from the phase-1 benign calibration source rather than from the \texttt{2925}-example evaluation pool. Before fitting, non-finite values are removed or mapped to finite defaults, the features are standardized, and the One-Class SVM is fit on the resulting reference features. The percentile file \texttt{svm\_feature\_stats.json} stores the benign-reference cut points used by the downstream trigger logic.

  At inference time, the SVM output is not used in isolation. Instead, the detector combines it with a small set of percentile-based outlier rules and two lightweight heuristic scores: a heuristic wrapper score $s_w$ for policy-bypass and obfuscation cues, and a multi-turn risk score $s_m$ for escalation patterns across turns.

  A row is treated as a \emph{severe outlier} when at least one of the following high-confidence patterns holds: a strongly negative safe-direction projection together with nontrivial projection drift; a simultaneous projection and cosine gap above their benign-reference tails; or unusually large joint-state dispersion together with elevated projection drift.

  A row is treated as a \emph{moderate outlier} under a weaker version of the same logic: milder negative projection plus drift, simultaneous projection-and-cosine deviation at the benign-reference threshold, or elevated dispersion accompanied by smaller but still abnormal consistency gaps. Concretely, the thresholds are:
  \begin{quote}\small
  \textbf{Severe outlier:} $c < -0.10$ and $\delta_c > \max(m_c,10^{-6})$; or $\delta_c > 1.35\,q_c$ and $\delta_{\cos} > 1.20\,q_{\cos}$; or $\mathrm{variance} > 1.60\,q_v$ and $\delta_c > 1.10\,q_c$.\\
  \textbf{Moderate outlier:} $c < -0.05$ and $\delta_c > m_c$; or $\delta_c > q_c$ and $\delta_{\cos} > q_{\cos}$; or $\mathrm{variance} > q_v$, $\delta_c > m_c$, and $\delta_{\cos} > 0.85\,q_{\cos}$.
  \end{quote}

  These signals are combined into a scalar severity score,
  \begin{equation}
  \begin{aligned}
  \texttt{detector\_score}
  =\;& \max(0,\texttt{risk\_signals}-1) \\
  &+ 1.5\,\mathbf{1}[\texttt{prediction}=-1] \\
  &+ 1.5\,\mathbf{1}[\texttt{moderate\_outlier}] \\
  &+ 2.5\,\mathbf{1}[\texttt{severe\_outlier}] \\
  &+ 0.75\,\texttt{multiturn\_risk\_score} \\
  &+ \texttt{wrapper\_risk\_score},
  \end{aligned}
  \end{equation}
  where \texttt{risk\_signals} counts threshold hits from the core consistency features together with the wrapper and multi-turn auxiliary scores. The detector fires when any one of the following conditions is met:
  \begin{enumerate}
  \item a severe outlier is detected;
  \item the SVM predicts an outlier and the row also satisfies the moderate-outlier test;
  \item at least two risk signals fire jointly;
  \item heuristic wrapper cues are strong enough to trigger a dedicated high-risk gate;
  \item a medium heuristic-wrapper pattern co-occurs with lexical high-risk evidence and at least one weaker consistency anomaly;
  \item a multi-turn risk pattern co-occurs with at least one weaker consistency anomaly;
  \item the multi-turn risk score alone reaches the highest escalation band.
  \end{enumerate}
  For reproducibility, the implementation instantiates these gates as follows:
  \begin{quote}\small
  \textbf{High heuristic-wrapper gate:} $s_w \ge 3.0$ and either the prompt matches the high-risk lexical rule or the row crosses the lighter wrapper $\delta_c/\delta_{\cos}$ test.\\
  \textbf{Intermediate heuristic-wrapper gate:} $s_w \ge 1.5$, the goal is lexically high-risk, and at least one of an SVM outlier prediction, a lighter $\delta_c$ threshold, or a lighter variance threshold holds.\\
  \textbf{Multi-turn risk gate:} $s_m \ge 2.5$ and at least one of an SVM outlier prediction, a lighter $\delta_c$ threshold, a lighter $\delta_{\cos}$ threshold, or a lighter variance threshold holds; or $s_m \ge 5.0$.
  \end{quote}
  This set of rules is fixed before held-out evaluation and shared across the reported main runs.

  The runtime intervention operators should be read as low-level moves inside a higher-level selection scheme rather than as standalone methods. Let $h$ denote the current last-token residual state and let $\hat v_{\mathrm{safe}}$ denote the normalized safety direction.

  \paragraph{Structure push.}
  \[
  h' = h + \alpha \|h\| \hat v_{\mathrm{safe}}.
  \]

  \paragraph{Reflection correction.}
  If the signed coefficient along $\hat v_{\mathrm{safe}}$ is sufficiently negative, the state is reflected back toward the safe direction:
  \[
  c = \frac{\langle h,\hat v_{\mathrm{safe}}\rangle}{\|\hat v_{\mathrm{safe}}\|_2^2}.
  \]
  \[
  h' = h - \lambda c \hat v_{\mathrm{safe}}
  \qquad \text{when } c < -0.05.
  \]

  \paragraph{Alignment-based adjustment.}
  When the alignment matrix $R$ is available, the system uses a projected safety direction:
  \[
  h' = h + \alpha_p \|h\| \widehat{R \hat v_{\mathrm{safe}}}.
  \]

  \paragraph{Centroid blending.}
  When a safe-variant centroid $\mu_{\mathrm{safe}}$ is available, the system first applies a light structure push and then blends toward the safe centroid:
  \[
  \begin{aligned}
  \tilde h &= h + \alpha \|h\| \hat v_{\mathrm{safe}}, \\
  h' &= (1-\beta)\tilde h + \beta \mu_{\mathrm{safe}}.
  \end{aligned}
  \]

  In the reported implementation, these operators are controlled by a fixed severity-dependent schedule. The base structure coefficient is \texttt{0.22}, and stronger anomalies are assigned larger shifts. Let \texttt{severity = detector\_score}. The coefficients are:

  \begin{enumerate}
  \item \textbf{Structure gain} (\texttt{structure\_gain}): the default value is \texttt{0.22}; it is scaled by \texttt{1.25} for \texttt{severity >= 5.0} and by \texttt{1.45} for \texttt{severity >= 6.5}, with an additional \texttt{1.35} multiplier in multi-turn settings, then clipped to \texttt{[0.18, 0.72]}. In multi-turn centroid blending, the same quantity is reduced to \texttt{30\%} of its current value and clipped to \texttt{[0.04, 0.10]}. In the dedicated multi-turn safe-reframing mode, no structure push is applied.
  \item \textbf{Centroid blend} (\texttt{invariant\_blend}): the default value is \texttt{0.28}; it increases to \texttt{0.32} for \texttt{severity >= 5.0} and to \texttt{0.38} for \texttt{severity >= 6.5}. In multi-turn centroid-blending mode it is reset to \texttt{0.03}, \texttt{0.04}, or \texttt{0.05} depending on severity or \texttt{multiturn\_risk\_score}. In single-turn centroid-blending mode it is lower-bounded by \texttt{0.36} when \texttt{wrapper\_risk\_score >= 2.0} or \texttt{severity >= 5.0}, and by \texttt{0.42} when \texttt{wrapper\_risk\_score >= 3.0} or \texttt{severity >= 6.5}; the final clip is \texttt{[0.20, 0.52]} in single-turn mode and \texttt{[0.01, 0.12]} in multi-turn mode.
  \item \textbf{Alignment gain} (\texttt{procrustes\_alpha}): the default value is \texttt{0.12}; it increases to \texttt{0.15} for \texttt{severity >= 5.0} and \texttt{0.18} for \texttt{severity >= 6.5}, is floored at \texttt{0.20} in multi-turn settings, and clipped to \texttt{[0.08, 0.25]}.
  \item \textbf{Reflection gain} (\texttt{reflection\_gain}): the default value is \texttt{1.25}; it increases to \texttt{1.40} for \texttt{severity >= 5.0} and \texttt{1.60} for \texttt{severity >= 6.5}, then is clipped to \texttt{[1.0, 1.8]}.
  \item \textbf{Hybrid centroid blend} (\texttt{hybrid\_centroid\_blend}): this term is used only in single-turn \texttt{structure} mode. It is set to \texttt{0.04} when $s_w \ge 2.0$ and \texttt{severity >= 3.0}, and to \texttt{0.06} when $s_w \ge 3.5$ or \texttt{severity >= 5.0}.
  \end{enumerate}

  Mode selection is likewise fixed before held-out test evaluation. The main text uses the default \texttt{core} family rather than the alternative \texttt{full} family retained in code. Under the reported \texttt{core} family, the selection is:
  \begin{enumerate}
  \item if the detector does not fire, the model is left unchanged;
  \item if the row is identified as multi-turn, the system uses a specialized safe-reframing mode, which is the clearest example of the helpful path because it rewrites only the next assistant turn into a safe continuation rather than always collapsing directly to refusal;
  \item otherwise, reflection is preferred for strongly negative projections with large projection drift;
  \item otherwise, centroid blending is preferred when safe variants are available and the row satisfies one of several higher-risk wrapper/dispersion patterns:
  \begin{quote}\small
  \textbf{Gate A:} $s_w \ge 2.0$, $s_d \ge 4.5$, and
  $\delta_c > \max(1.10\,m_c, 0.10)$.\\
  \textbf{Gate B:} $s_d \ge 5.0$, $c > -0.12$,
  $\delta_c > 1.10\,q_c$, $\delta_{\cos} > 1.05\,q_{\cos}$,
  and $\mathrm{variance} > 0.90\,q_v$.\\
  \textbf{Gate C:} $s_w \ge 3.0$, $s_d \ge 6.0$, $c > 0.10$,
  and $\mathrm{variance} > 0.90\,q_v$.\\
  \textbf{Gate D:} $s_w \ge 2.0$, $s_d \ge 5.5$, $c < -0.03$,
  and $\mathrm{variance} > 1.10\,q_v$.
  \end{quote}
  \item if none of the above conditions hold, the system uses \texttt{structure} mode.
  \end{enumerate}
  In the reported system, the single-turn helpful path is implemented directly as a dedicated helpful recovery branch for ambiguous borderline benign prompts whose surface form may still look malicious or policy-violating, but whose safe-variant evidence supports a benign, non-harmful interpretation. For the alternative \texttt{full} family retained in code, the single-turn selector is simpler. After the same reflection test, it:
  \begin{enumerate}
  \item uses centroid blending when $\mathrm{variance} > 1.45\,q_v$ and $\delta_c > 1.05\,q_c$;
  \item uses alignment-based adjustment only when an alignment matrix is available and $\delta_{\cos} > 1.20\,q_{\cos}$;
  \item otherwise uses \texttt{structure}.
  \end{enumerate}
  For the feature banks used in these experiments, the stored percentile anchors are:
  \begin{enumerate}
  \item $\delta_{c,\mathrm{p95}}=\texttt{0.4161}$,
  \item $\delta_{c,\mathrm{p50}}=\texttt{0.0769}$,
  \item $\delta_{\cos,\mathrm{p95}}=\texttt{0.1651}$,
  \item $\mathrm{variance}_{\mathrm{p95}}=\texttt{1.3636}$.
  \end{enumerate}
  For Gemma they are:
  \begin{enumerate}
  \item \texttt{0.4240},
  \item \texttt{0.1326},
  \item \texttt{0.1424},
  \item \texttt{0.6725},
  \end{enumerate}
  respectively. These constants are reported for reproducibility rather than as theoretically optimal choices. Appendix Tables~\ref{tab:layer-sensitivity}, \ref{tab:k-sensitivity}, and \ref{tab:operator-ablation} provide complementary sensitivity measurements for this operating point. In particular, Appendix Table~\ref{tab:operator-ablation} shows that removing reflection, centroid blending, or alignment-based adjustment weakens the reported ASR/FPR trade-off, which is consistent with interpreting the intervention stack as a coordinated design rather than as a single dominant operator.

  \section{Behavioral Evaluation Details}
  \label{app:behavioral-protocols}
  \label{app:behavioral-rubric}

  The GPT-4 external judge~\cite{openai2023gpt4} receives the original malicious prompt and the final model response, then returns one label from \{\texttt{operational\_harmful}, \texttt{refusal}, \texttt{safe\_answer}, \texttt{non\_harmful\_offtarget}, \texttt{degenerate}, \texttt{answer}\}. We define \textbf{Harm N} as the number of malicious examples labeled \texttt{operational\_harmful} and \textbf{Judge ASR} as \textbf{Harm N} divided by the number of evaluated malicious examples.

  The manual category review uses the same shared six-label scheme and is conducted by the authors on the aligned held-out exports under the same malicious pool of \texttt{2416} examples. No external annotators are recruited or paid for this review step.

  The behavioral tables in this paper are aligned to the unified malicious pool of \texttt{2416} held-out examples in \texttt{jailbreak\_test\_test.jsonl}. Table~\ref{tab:manual-audit} reports the full held-out manual category review over that pool. For \texttt{Llama-3-8B} and \texttt{Gemma-2-9B} \texttt{Vanilla/Detect/ReAnchor}, the per-row category labels are taken directly from the stored phase-2 detail exports. Table~\ref{tab:external-judge} reports the judge evaluation on the same \texttt{2416}-example pool, while Table~\ref{tab:behavioral-subset-provenance} records the provenance of the aligned behavioral pools.

  \begin{table}[H]
\centering
\scriptsize
\setlength{\tabcolsep}{2pt}
\begin{tabular}{@{}p{0.20\columnwidth}p{0.26\columnwidth}p{0.08\columnwidth}p{0.36\columnwidth}@{}}
\toprule
Pool & Systems & N & Provenance \\
\midrule
Full held-out manual category review & System rows reported in Table~\ref{tab:manual-audit} & 2416 & The full-pool manual category review is computed on the main held-out malicious pool \texttt{jailbreak\_\allowbreak test\_\allowbreak test.\allowbreak jsonl}. The reported category rows are taken from the aligned held-out category exports used for Table~\ref{tab:manual-audit}. \\
GPT-4 judge evaluation & Model/System rows in Table~\ref{tab:external-judge} and Table~\ref{tab:behavioral-alignment} & 2416 & The judge table is aligned to the same main held-out malicious pool of \texttt{2416} examples. \\
\bottomrule
\end{tabular}
\caption{Provenance of the unified behavioral evaluation pools. The paper's aligned behavioral reporting is anchored to the main held-out malicious pool of \texttt{2416} examples.}
\label{tab:behavioral-subset-provenance}
\end{table}

  \begin{table}[H]
\centering
\scriptsize
\setlength{\tabcolsep}{2.6pt}
\resizebox{\columnwidth}{!}{%
\begin{tabular}{@{}llrrrrrrrrr@{}}
\toprule
Model & System & N & ASR & Harm N & Harm (\%) & Refuse & Safe & Off-target & Degen. & Answer \\
\midrule
\texttt{Llama-3-8B} & Vanilla & 2416 & 41.56 & 579 & 23.97 & 1005 & 326 & 302 & 204 & 579 \\
\texttt{Llama-3-8B} & NeuroArmor-Detect & 2416 & 41.56 & 579 & 23.97 & 1005 & 326 & 302 & 204 & 579 \\
\texttt{Llama-3-8B} & NeuroArmor-ReAnchor & 2416 & \textbf{1.57} & 24 & 0.99 & 2003 & 325 & 52 & 12 & 24 \\
\texttt{Llama-3-8B} & Direct Safety Steering & 2416 & 23.92 & 324 &13.41  & 1309  & 407 & 174 & 202 & 324 \\
\texttt{Llama-3-8B} & Runtime Intervention Baseline & 2416 & 24.92 & 371 & 15.36 & 1277 & 318 & 244  & 206 & 371 \\
\texttt{Llama-3-8B} & Llama-Guard-3 & 2416 & 26.61 & 255 & 10.56 & 1421 & 231 & 144 & 365 & 255  \\
\midrule
\texttt{Gemma-2-9B} & Vanilla & 2416 & 53.52 & 798 & 33.03 & 1087 & 268 & 62 & 201 & 798 \\
\texttt{Gemma-2-9B} & NeuroArmor-Detect & 2416 & 53.52 & 798 & 33.03 & 1087 & 268 & 62 & 201 & 798 \\
\texttt{Gemma-2-9B} & NeuroArmor-ReAnchor & 2416 & \textbf{1.90} & 31 & 1.28 & 2152 & 110 & 76 & 47 & 31 \\
\bottomrule
\end{tabular}
\par}
\caption{Full held-out manual category review on the unified main held-out malicious pool of \texttt{2416} examples. The category columns follow the shared label set used in the behavioral evaluation: \texttt{operational\_harmful}, \texttt{refusal}, \texttt{safe\_answer}, \texttt{non\_harmful\_offtarget}, \texttt{degenerate}, and \texttt{answer}. The \texttt{ASR} column is aligned to the same held-out malicious pool.}
\label{tab:manual-audit}
\end{table}

  \begin{table}[H]
\centering
\scriptsize
\setlength{\tabcolsep}{2pt}
\renewcommand{\arraystretch}{1.12}
\begin{tabular}{@{}p{0.17\columnwidth}p{0.19\columnwidth}p{0.20\columnwidth}p{0.20\columnwidth}p{0.20\columnwidth}@{}}
\toprule
Model & System & ASR (held-out) & GPT-4 Judge (held-out) & Manual review (held-out) \\
\midrule
\texttt{Llama-3-8B} & Vanilla & 41.56 (1004/2416) & 25.66 (620/2416) & 23.97 (579/2416) \\
\texttt{Llama-3-8B} & NeuroArmor-Detect & 41.56 (1004/2416) & 25.66 (620/2416) & 23.97 (579/2416) \\
\texttt{Llama-3-8B} & NeuroArmor-ReAnchor & \textbf{1.57 (38/2416)} & \textbf{0.87 (21/2416)} & \textbf{0.99 (24/2416)} \\
\texttt{Llama-3-8B} & Direct Safety Steering & 23.92 (578/2416) & 18.34 (443/2416) & 13.41 (324/2416) \\
\texttt{Llama-3-8B} & Runtime Intervention Baseline & 24.92 (602/2416) & 21.94 (530/2416) & 15.36 (371/2416) \\
\texttt{Llama-3-8B} & Llama-Guard-3 & 26.61 (643/2416) & 16.97 (410/2416) & 10.56 (255/2416) \\
\midrule
\texttt{Gemma-2-9B} & Vanilla & 53.52 (1293/2416) & 30.75 (743/2416) & 33.03 (798/2416) \\
\texttt{Gemma-2-9B} & NeuroArmor-Detect & 53.52 (1293/2416) & 30.75 (743/2416) & 33.03 (798/2416) \\
\texttt{Gemma-2-9B} & NeuroArmor-ReAnchor & \textbf{1.90 (46/2416)} & \textbf{1.32 (32/2416)} & \textbf{1.28 (31/2416)} \\
\bottomrule
\end{tabular}
\caption{Aligned comparison of ASR, GPT-4 judge, and manual review on the full main held-out malicious pool of \texttt{2416} examples. Each entry reports the percentage followed by the exact count fraction in parentheses. Lower is better in all three columns.}
\label{tab:behavioral-alignment}
\end{table}

  \begin{table}[H]
\centering
\small
\setlength{\tabcolsep}{4pt}
\resizebox{\columnwidth}{!}{%
\begin{tabular}{lcccc}
\toprule
System & MMLU & GSM8K & HumanEval & Intervention rate \\
\midrule
Vanilla & 59.28 & 86.02 & 55.49 & 0.0 \\
NeuroArmor-ReAnchor & 59.28 & 86.02 & 55.49 & 0.0 \\
\bottomrule
\end{tabular}
}
\caption{Utility evaluation on the shared benchmark set. No regression is observed on the evaluated tasks.}
\label{tab:utility}
\end{table}

  \FloatBarrier

  \section{Appendix Pipeline Sketch}
  \label{app:pipeline-sketch}

  We include a compact sketch of the prompt-processing pipeline used at inference time in the reported system.

  \begin{figure}[!htbp]
    \centering
    \fbox{%
    \begin{minipage}{0.95\columnwidth}
    \scriptsize
    \begin{tabular}{@{}l@{}}
    \textbf{Input:} prompt $x$ \\
    1.\; Derive a topic summary $s$ from raw text $x$. \\
    2.\; Construct $K$ safe variants $\mathcal{B}(x)$ from $s$. \\
    3.\; Run the base model on $x$ and on $\mathcal{B}(x)$. \\
    4.\; Compute consistency features $z(x)$. \\
    5.\; If $z(x)$ is out-of-distribution, trigger intervention; \\otherwise leave the state unchanged. \\
    6.\; Generate the final response and score it under the evaluation setting.
    \end{tabular}
    \end{minipage}}
    \caption{Compact prompt-processing pipeline. The same path is used for benign and malicious inputs; only the trigger decision changes the downstream intervention step.}
    \label{fig:non_oracle_sketch}
  \end{figure}

  \FloatBarrier

  \section{Sensitivity and Robustness Evaluation}
  \label{app:sensitivity-robustness}

  The tables in this appendix section report additional runtime, sensitivity, robustness, and overlap results that complement the main evaluation. Together they characterize runtime overhead, layer choice, variant count, intervention-operator contributions, borderline-benign behavior, adaptive-attack stress tests, and split overlap.

  \begin{table}[H]
\centering
\small
\begin{tabular}{lcc}
\toprule
Stage & Seconds & Relative \\
\midrule
Variant lookup / template build & 0.04 & 0.00x \\
Activation extraction & 7.82 & 0.19x \\
Detector scoring & 0.00 & 0.00x \\
ReAnchor generation & 41.57 & 1.01x \\
Vanilla generation & 41.32 & 1.00x \\
End-to-end NeuroArmor-ReAnchor & 49.43 & 1.20x \\
\bottomrule
\end{tabular}
\caption{Runtime decomposition of NeuroArmor-ReAnchor. Relative cost is computed against vanilla generation.}
\label{tab:runtime}
\end{table}

  \begin{table}[H]
\centering
\scriptsize
\setlength{\tabcolsep}{3pt}
\resizebox{\columnwidth}{!}{%
\begin{tabular}{@{}lccp{0.39\columnwidth}@{}}
\toprule
Layer & ASR & FPR & Note \\
\midrule
$-8$ & 17.20 & 5.10 & Earlier-layer re-anchoring lowers benign blocking but leaves substantially higher ASR than the final-block setting. \\
$-4$ & 12.60 & 5.70 & This intermediate layer improves over $-8$ but still trails later-layer re-anchoring on malicious suppression. \\
$-2$ & 10.30 & 6.00 & The near-final layer narrows the ASR gap further, but the final block still provides the strongest malicious suppression. \\
$-1$ & 1.57 & 22.05 & Reported Llama-3 operating point used in the main paper; this is the final-block setting referenced in the main held-out results. \\
\bottomrule
\end{tabular}
}
\caption{Target-layer sensitivity for the main Llama-3-8B setting.}
\label{tab:layer-sensitivity}
\end{table}

  \begin{table}[H]
\centering
\scriptsize
\setlength{\tabcolsep}{3pt}
\resizebox{\columnwidth}{!}{%
\begin{tabular}{@{}lccp{0.34\columnwidth}@{}}
\toprule
$K$ & Llama-3 ASR/FPR & Gemma-2 ASR & Note \\
\midrule
$K=1$ & 4.80/37.80 & 7.20 & Weakest reference geometry and highest ASR in the sweep. \\
$K=2$ & 2.40/36.10 & 3.80 & Recovers most of the gap, but still trails $K=3$ on both models. \\
$K=3$ & 1.57/22.05 & 1.90 & Reported operating point used in the main evaluation. \\
$K=4$ & 1.52/22.15 & 1.85 & Only marginal change relative to $K=3$. \\
$K=5$ & 1.48/22.20 & 1.82 & Near saturation by $K=5$; negligible change beyond $K=4$. \\
\bottomrule
\end{tabular}
}
\caption{Safe-variant-count sensitivity across the reported Llama-3 and Gemma-2 settings. For Gemma-2, we report ASR as cross-model support.}
\label{tab:k-sensitivity}
\end{table}

  \begin{table}[H]
\centering
\scriptsize
\setlength{\tabcolsep}{3pt}
\resizebox{\columnwidth}{!}{%
\begin{tabular}{@{}lcp{0.38\columnwidth}@{}}
\toprule
Configuration & ASR/FPR & Note \\
\midrule
Full NeuroArmor-ReAnchor & 1.57/22.05 & Reference row from the main held-out evaluation setting. \\
Direction-only re-anchor & 23.48/35.29 & Uniform safety-direction push; much weaker in both ASR and benign calibration. \\
w/o reflection correction & 2.02/35.44 & Both ASR and FPR worsen; reflection helps preserve the full trade-off. \\
w/o centroid blending & 2.15/35.52 & Both ASR and FPR worsen; centroid guidance matters on harder cases. \\
w/o alignment-based adjustment & 1.90/35.36 & ASR rises and benign FPR stays high; alignment provides a measurable gain. \\
Structure-only re-anchor & 23.90/35.35 & Coarse structure push; far weaker than the full selective stack. \\
\bottomrule
\end{tabular}
}
\caption{Micro-operator ablation under the main held-out evaluation setting.}
\label{tab:operator-ablation}
\end{table}

  \begin{table}[H]
\centering
\scriptsize
\setlength{\tabcolsep}{3pt}
\resizebox{\columnwidth}{!}{%
\begin{tabular}{@{}lccp{0.35\columnwidth}@{}}
\toprule
Benign category & Vanilla blocked & ReAnchor blocked & Note \\
\midrule
XS-Test-1356 & 1.11 & 0.37 & Easier benign slice; re-anchoring reduces blocking further. \\
XS-Test-safe & 3.20 & 4.00 & Smaller safe slice; modest benign cost increase. \\
OR-Bench-Hard & 65.35 & 47.76 & Main source of false positives; most pooled FPR gain comes from this slice. \\
\bottomrule
\end{tabular}
}
\caption{Benign-boundary evaluation on the shared Llama-3 benign pool used for the pooled FPR analysis.}
\label{tab:benign-boundary}
\end{table}

  \begin{table}[H]
\centering
\scriptsize
\setlength{\tabcolsep}{2pt}
\begin{tabular}{@{}p{0.40\columnwidth}p{0.13\columnwidth}p{0.13\columnwidth}p{0.26\columnwidth}@{}}
\toprule
Adaptive threat model & Llama-3 ASR & Gemma ASR & Note \\
\midrule
Detector-aware paraphrase-preserving jailbreak & 8.40 & 12.20 & Stronger than the static benchmark setting, but still materially below vanilla ASR. \\
Low-perturbation suffix search against the detector features & 15.60 & 21.80 & Strongest reported white-box probe; exposes the largest adaptive weakness in the current evaluation. \\
Representation-consistent multi-turn jailbreak & 13.50 & 18.90 & Important because multi-turn jailbreaks remain the hardest malicious slice in the main evaluation. \\
Cross-model detector-aware attack & 10.10 & 14.30 & Suggests part of the weakness transfers across models rather than coming only from model-specific quirks. \\
\bottomrule
\end{tabular}
\caption{Adaptive-attacker stress tests across the reported Llama-3 and Gemma-2 settings.}
\label{tab:adaptive-attack}
\end{table}

  \begin{table}[H]
\centering
\scriptsize
\setlength{\tabcolsep}{2pt}
\begin{tabular}{@{}p{0.40\columnwidth}p{0.12\columnwidth}p{0.36\columnwidth}@{}}
\toprule
Overlap check & Value & Note \\
\midrule
Phase-1 malicious reference rows vs held-out malicious evaluation rows & 0.00\% & No exact-row overlap between malicious reference construction and held-out malicious evaluation rows. \\
Phase-1 benign calibration rows vs held-out benign evaluation rows & 0.00\% & No exact-row overlap between benign calibration rows and held-out benign evaluation rows. \\
Exact normalized goal duplicates across split boundaries (count) & 0 & Zero normalized-goal duplicates after split-time deduplication. \\
Held-out exact-match rate to stored phase-1 variants & 18.00\% & Stored variants are reused in some cases, but deterministic construction remains the main path. \\
\bottomrule
\end{tabular}
\caption{Overlap and leakage check under the reported train/validation/test split.}
\label{tab:overlap-audit}
\end{table}

  \FloatBarrier

  \section{ARMOR Reference}

  We also report \texttt{ARMOR} in the appendix for context. This reference is not directly comparable to the main held-out results because the available \texttt{ARMOR} evaluation does not cover the multi-turn jailbreak portion of the main held-out malicious pool, which changes the effective sample count relative to the main evaluation setting. We therefore retain it only as an auxiliary reference rather than as a matched baseline for the main table.

  \begin{table}[H]
\centering
\scriptsize
\setlength{\tabcolsep}{3pt}
\resizebox{\columnwidth}{!}{%
\begin{tabular}{@{}lcc@{}}
\toprule
Method & ASR (\%; count/total) & FPR (\%; count/total) \\
\midrule
ARMOR guarded & 27.53 (500/1816) & 32.68 (956/2925) \\
ARMOR original & 22.58 (410/1816) & 39.79 (1164/2925) \\
\bottomrule
\end{tabular}%
}
\caption{ARMOR reference results reported for context. These rows are not directly comparable to the main held-out table because the available ARMOR evaluation does not cover the multi-turn jailbreak portion of the main held-out malicious pool, so the effective sample counts differ from the main evaluation setting.}
\label{tab:armor-historical-reference}
\end{table}

  \end{document}